\documentclass[aps,prd,preprint,groupedaddress,amssymb,nofootinbib]{revtex4}

\usepackage{color}
\usepackage{epsfig}
\usepackage{graphicx}
\usepackage{epstopdf}

\def\be{\begin{equation}}
\def\ee{\end{equation}}

\def\ifmath#1{\relax\ifmmode #1\else $#1$\fi}
\def\half{\ifmath{{\textstyle{1 \over 2}}}}

\def\mh{m_h}

\def\mai{m_{a_1}}

\def\br{{\rm Br}}
\def\tanb{\tan\beta}

\def\mhusq{m_{H_u}^2}
\def\mhdsq{m_{H_d}^2}

\def\hsm{h_{\rm SM}}
\def\mhsm{m_{\hsm}}

\def\tanb{\tan\beta}

\def\mb{m_b}
\def\mz{M_Z}

\def\mgut{M_{\rm GUT}}

\def\eg{{\it e.g.}}

\newcommand{\nc}{\newcommand}
\nc{\beq}{\begin{equation}}   \nc{\eeq}{\end{equation}}
\nc{\bea}{\begin{eqnarray}}   \nc{\eea}{\end{eqnarray}}
\nc{\baa}{\begin{array}}      \nc{\eaa}{\end{array}}
\nc{\bit}{\begin{itemize}}    \nc{\eit}{\end{itemize}}
\nc{\ben}{\begin{enumerate}}  \nc{\een}{\end{enumerate}}
\nc{\bce}{\begin{center}}     \nc{\ece}{\end{center}}
\def\beqa{\begin{eqnarray}}
\def\eeqa{\end{eqnarray}}  
\def\bed{\begin{description}}
\def\eed{\end{description}}

\def\gev{~{\rm GeV}}

\def\tev{~{\rm TeV}}

\def\hsm{h_{\rm SM}}
\def\mhsm{m_{\hsm}}

\newcommand\lsim{\mathrel{\rlap{\lower4pt\hbox{\hskip1pt$\sim$}}
    \raise1pt\hbox{$<$}}}
\newcommand\gsim{\mathrel{\rlap{\lower4pt\hbox{\hskip1pt$\sim$}}
    \raise1pt\hbox{$>$}}}
\def\bea{\begin{eqnarray}}
\def\eea{\end{eqnarray}}
\def\ba{\begin{array}}
\def\ea{\end{array}}
\def\bc{\begin{center}}
\def\ec{\end{center}}

\def\f{\frac}

\def\f#1#2{\frac{#1}{#2}}

\begin{document}

\preprint{IUHET-529}
\title{Unusual Higgs or Supersymmetry\\
from Natural Electroweak Symmetry Breaking\footnote{This review is based on many seminars and conference talks given by the author in 2005-2008.}}

\author{Radovan Derm\' \i\v sek}

\affiliation{Physics Department, Indiana University, Bloomington, IN 47405, USA}

\date{July 22, 2009}


\begin{abstract}
This review provides an elementary discussion of electroweak symmetry breaking in the minimal and the next-to-minimal supersymmetric models 
with the focus on the fine-tuning problem -- the tension between natural electroweak  symmetry breaking and the direct search limit on the Higgs boson mass.
Two generic solutions of the fine-tuning problem are discussed in detail: 
 models with unusual Higgs decays; 
and  models with unusual pattern of soft supersymmetry breaking parameters.
\end{abstract}

\maketitle

\section{Introduction}

Minimal Supersymmetric Standard Model  (MSSM) is a promising candidate
for describing physics above the electroweak (EW) scale. The three gauge
couplings unify~\cite{Dimopoulos:1981yj,Dimopoulos:1981zb,Ibanez:1981yh,Sakai:1981gr} at the  grand unified theory (GUT) scale $\sim~2~\times~10^{16}$~GeV
within a few percent, and 
the hierarchy between the EW scale and the GUT scale is naturally stabilized by
supersymmetry (SUSY)~\cite{Witten:1981nf}. 
In addition, if we add soft-supersymmetry-breaking terms (SSBs) at the GUT scale
we typically find that the mass squared of the Higgs doublet which couples to the 
top quark ($H_u$), is driven to negative values at the EW scale. This triggers 
electroweak symmetry breaking~\cite{Ibanez:1982fr} and the EW scale is naturally understood from 
SUSY breaking scale.
Furthermore, assuming R-parity, the lightest supersymmetric
particle (LSP) is stable and it is a natural candidate for
dark matter of the universe.

The real virtue of supersymmetry is that the above mentioned features do not 
require any specific relations between soft-supersymmetry-breaking parameters 
and the only strong requirement on SUSY breaking scenarios is that these terms 
are of order  the EW scale. However generic SSBs near the EW scale generically 
predict too light Higgs mass which is ruled out by LEP limits. The exact value of 
the Higgs mass is not relevant for low energy physics, nothing crucially depends 
on it, and yet, in order to stay above LEP limits ($m_h \gsim 114.4$ 
GeV)~~\cite{Barate:2003sz} the SSBs 
have to be either considerably above the EW scale and related to each other (or to the $\mu$ term) in a 
non-trivial way. SSBs can no longer be just generic which leads to strong 
requirements 
on possible models for SUSY breaking should these provide a natural explanation for 
the scale at which the electroweak symmetry is broken. 

In this review we discuss in detail the electroweak symmetry breaking in the minimal and the next-to-minimal supersymmetric models 
and focus on the fine-tuning problem -- the tension between natural electroweak  symmetry breaking and the direct search limit on the Higgs boson mass.
We will assume that the low scale supersymmetry is the scenario responsible for EWSB, and more importantly, the EWSB happens in a natural way without necessity of fine tuning.\footnote{Of course it is possible that idea of low scale supersymmetry is wrong and there is a different mechanism realized in nature that is responsible for EWSB. For a review of ideas, see {\it e.g.} Ref.~\cite{Giudice:2007qj}.} We will see that this assumption will lead us to consider models in which the Higgs boson decays in an unusual way or models 
with unusual pattern of soft supersymmetry breaking parameters.

\section{Electroweak Symmetry Breaking in the MSSM, the Higgs Mass and the Fine Tuning Problem}

Let us start with reviewing reasons we expect superpartners at the EW scale. Confronting these expectations with constraints imposed by limits on the Higgs mass  will lead us to  the fine-tuning problem.

\subsection{Electroweak symmetry breaking in the MSSM}

The mass of the $Z$ boson (or the EW scale), determined by
minimizing the Higgs potential, is related to the supersymmetric
Higgs mass parameter $\mu$ and the soft-SUSY-breaking mass-squared 
parameters for $H_u$  and $H_d$ by:
\beq
\half\mz^2=-\mu^2+{\mhdsq(M_Z) -  \mhusq(M_Z) \tan^2\beta  \over \tan^2\beta-1}\,.
\label{mzsquared}
\eeq
where $\tan \beta = v_u/v_d$ is the ratio of  vacuum expectation values 
of $H_u$ and $H_d$.
For $\tan \beta$ larger than $\gtrsim 5$ the formula simplifies to
\be 
\half M_Z^2
\simeq  -\mu^2 (M_Z) - m_{H_u}^2 (M_Z). 
\label{eq:MZ} 
\ee 
The EW scale value of $m_{H_u}^2$ depends on the boundary condition at a high scale and the correction accumulated through the 
renormalization group (RG) evolution:
\be 
m_{H_u}^2 (M_Z) = m_{H_u}^2 + \; \delta m_{H_u}^2,
\label{eq:mHu2} 
\ee 
and thus we find
\be 
\half M_Z^2
\simeq  -\mu^2 (M_Z) - m_{H_u}^2 -  \; \delta m_{H_u}^2.
\label{eq:MZdel} 
\ee 
The RG evolution of $\mu$ is controlled only by gauge and Yukawa couplings while the  
$\delta m_{H_u}^2$ depends on  
all soft-SUSY-breaking parameters. 
 For given $\tan \beta$ (which sets values of Yukawa couplings), we can solve the 
RG equations and express EW scale values of $m_{H_u}^2$,
$\mu^2$, and consequently $M_Z^2$ given by Eq. (\ref{eq:MZ}), in terms
of all GUT-scale parameters~\cite{Ibanez:1983di,Carena:1996km} (we consider the GUT-scale as an example, the conclusions 
do not depend on this choice). For $\tan \beta =10$, we
have:
\bea M_Z^2 & \simeq & -1.9 \mu^2 + 5.9 M_3^2 -1.2 m_{H_u}^2 + 1.5 m_{\tilde{t}}^2
  - 0.8 A_t M_3 + 0.2 A_t^2  + \cdots,
\label{eq:MZ_gut} 
\eea 
where parameters appearing on the right-hand side are the GUT-scale
parameters (we do not write the scale explicitly). Here, $M_3$ is the
$SU(3)$  gaugino mass, $A_t$ is the top soft-SUSY-breaking trilinear coupling,  and for simplicity we have defined $m^2_{\tilde{t}} \equiv (
m_{\tilde{t}_L}^2 + m_{\tilde{t}_R}^2)/2$ and assumed that  the
soft-SUSY-breaking stop mass-squared parameters are comparable, $m_{\tilde{t}_L}^2 \simeq m_{\tilde{t}_R}^2$.
Other scalar masses
and the $SU(1)$ and $SU(2)$ gaugino masses, $M_1$ and $M_2$, appear
with negligible coefficients and we neglect them in our discussion.
The coefficients in this expression depend only on $\tan \beta$ (they
do not change dramatically when varying $\tan \beta$ between 5 and 50)
and $\log (\mgut/\mz)$.  In a similar way we can express the EW-scale
values of the stop mass-squared, gluino mass and top trilinear
coupling in terms of GUT-scale boundary conditions. For $\tan \beta = 10$ we have:
\bea
m_{\tilde{t}}^2 (M_Z) & \simeq & 5.0 M_3^2 + 0.6 m_{\tilde{t}}^2   + 0.2 A_t M_3 \label{eq:mstop_gut} \\
M_3 (M_Z) & \simeq & 3 M_3 \label{eq:M3_gut} \\
A_t (M_Z) & \simeq & - 2.3 M_3 + 0.2 A_t. \label{eq:At_gut} 
\eea
Comparing Eq.~(\ref{eq:MZ_gut}) with Eqs.~(\ref{eq:mstop_gut}) and
(\ref{eq:M3_gut}) we easily see the usual expectation from SUSY:
\be
M_Z \simeq m_{\tilde{t}_{1,2}} \simeq m_{\tilde{g}} \simeq \mu,
\label{eq:expectation}
\ee 
when all the soft-SUSY-breaking
parameters are comparable. Furthermore, neglecting
terms proportional to $A_t$ in Eqs.~(\ref{eq:At_gut}) and
(\ref{eq:mstop_gut}) we find that the {\it typical stop mixing} is 
\be
\frac{| A_t |}{m_{\tilde{t}}} (M_Z)  \simeq  \frac{2.3
M_3}{\sqrt{5.0 M_3^2 + 0.6 m_{\tilde{t}}^2}} \lsim 1.0,
\ee 
which has an important implication for the Higgs mass.

\subsection{Typical Higgs mass}

The minimal supersymmetric model (MSSM) contains two Higgs doublets which result in five Higgs bosons in the spectrum:  light and heavy CP even Higgses, $h$ and $H$, the CP odd Higgs, $A$, and a pair of charged Higgs bosons, $H^\pm$. In the decoupling limit, $m_A \gg m_Z$, we find that $m_A \simeq m_H \simeq m_{H^\pm}$ and the light CP even Higgs has the same coupling to the Z-boson as the Higgs in the standard model; we say the light CP even Higgs boson is standard-model-like (SM-like). Its mass is approximately given as: 
\be
m_h^2 \simeq M_Z^2 \cos^2 2\beta + \frac{3G_F m_t^4}{\sqrt{2} \pi^2}
\left\{ \log \frac{m_{\tilde{t}}^2}{m_t^2} +
  \frac{A_t^2}{m_{\tilde{t}}^2} (1-\frac{A_t^2}{12 m_{\tilde{t}}^2} )
\right\}. \label{eq:mh_mix} 
\ee 
where the first term is the tree level result and the second term is
the dominant one-loop correction~\cite{Okada:1990vk,Haber:1990aw,Ellis:1990nz,Ellis:1991zd}.

\begin{figure}[ht]
 \centerline{\psfig{file=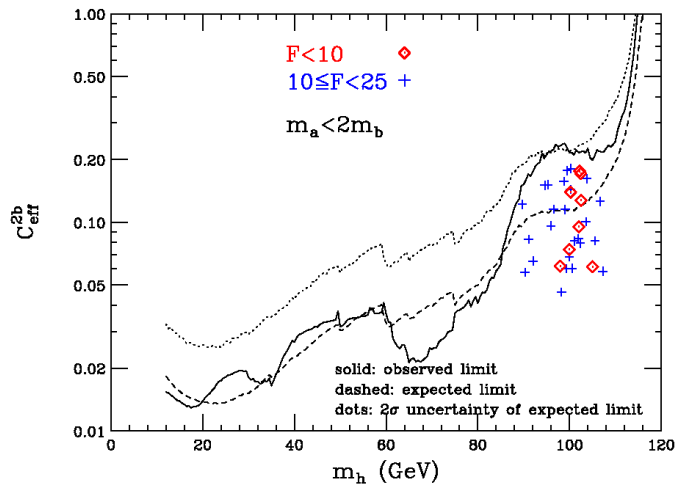,width=3.5in}}
\vspace*{8pt}
\caption{Expected and observed 95\% CL limits on $C^{2b}_{eff}$ 
are shown vs. $\mh$.  Also plotted are the
predictions for the NMSSM parameter cases 
having fixed $\tanb=10$, $M_{1,2,3}(\mz)=100,200,300\gev$ that give
 fine-tuning measure $F<25$ and $\mai<2\mb$ and that are consistent
  with Higgs constraints.}
\label{zbblimits}
\end{figure}

In the SM the dominant decay mode of the Higgs boson 
is $\hsm \to b \bar b$ when $\mhsm\lsim 140\gev$. This is also the dominant decay mode of the SM-like Higgs in most of the SUSY parameter space in the MSSM.
At LEP the SM-like Higgs boson could be produce in association with the Z-boson and
LEP has placed strong constraints on $Zh\to Z b\bar b$.  The
limits on
\beq
C^{2b}_{eff} \equiv [g_{ZZh}^2/g_{ZZh_{SM}}^2]\br(h\to b\bar b),
\eeq 
where $g_{ZZh}$ is the $Z-Z-h$ coupling in a given model, are shown in Fig.~\ref{zbblimits} (from
Ref.~\cite{Barate:2003sz,Dermisek:2007yt}). From this plot, one concludes that
$\mh<114\gev$ is excluded for a SM-like $h$ that decays primarily to
$b\bar b$.\footnote{For
  discussion of the possibility that $H$ is SM-like or that $h$ and
  $H$ share the coupling to $ZZ$ and $WW$ see {\it
    e.g.}~\cite{Dermisek:2007ah} and references therein. This possibility does not have a significant impact on the discussion of naturalness of EWSB which is the focus of this review.}

\begin{figure}[th]
\centerline{\psfig{file=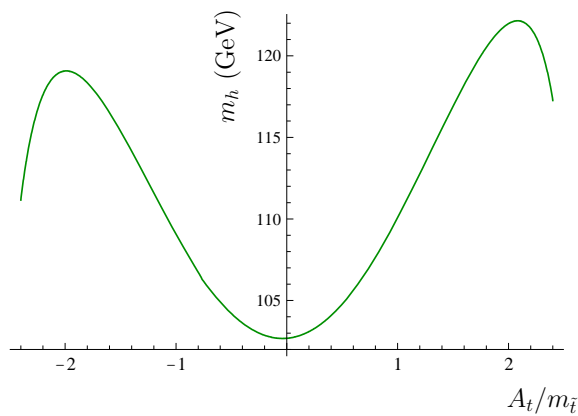,width=3.0in}}
\vspace*{8pt}
\caption{Mass of the Higgs boson in the MSSM as a function of the mixing in the stop sector, $ A_t /  m_{\tilde t} $, 
for $\tan \beta = 10$ and $m_{\tilde t} = 400$ GeV ($M_{SUSY} = m_A = \mu = 400$ GeV) obtained from {\it FeynHiggs-2.5.1}.
\label{fig:higgs_mass_vs_mixing}}
\end{figure}

\begin{figure}[th]
\centerline{\psfig{file=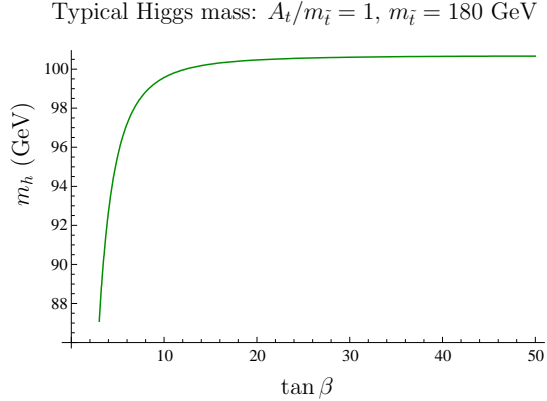,width=3.0in}}
\vspace*{8pt}
\caption{
Mass of the Higgs boson in the MSSM as a function of $\tan \beta$ for 
typical mixing in the stop sector, $- A_t /  m_{\tilde t}  = 1$, with 
$M_{SUSY} = m_A = \mu = 200$ GeV obtained from {\it FeynHiggs-2.5.1}.
\label{fig:typical_higgs_mass}}
\end{figure}

From, Eq.~(\ref{eq:mh_mix}) we see that at tree level, $m_h < M_Z \simeq 91$
GeV which does not satisfy the LEP limit. The Higgs mass can be increased beyond this value either by increasing the
mixing in the stop sector, $A_t/m_{\tilde{t}}$, or by increasing the
stop mass, $m_{\tilde{t}}^2$. The effect of stop mixing on the Higgs mass is shown in Fig.~\ref{fig:higgs_mass_vs_mixing} 
obtained from {\it FeynHiggs-2.5.1}~\cite{Heinemeyer:1998yj,Heinemeyer:1998np}.
As we have learned, the typical mixing
in the stop sector, which is achieved as a result of RG evolution from a large
range of high scale boundary conditions, is $|A_t|/m_{\tilde{t}}(M_Z)
\lsim 1.0$.  With this typical mixing and superpartners near the EW scale, we obtain the {\it typical Higgs
mass}, $m_h \simeq 100$ GeV, see Fig.~\ref{fig:typical_higgs_mass}. 
This is a prediction from a large range of SUSY parameter space and it is highly insensitive to
small variations of soft SUSY breaking terms at the unification scale (for fixed mixing the Higgs mass depends only logarithmically on stop masses).

In order to push the Higgs mass above the
LEP limit, 114.4 GeV, assuming the typical mixing, the stop masses
have to be $\gsim 1\tev$.\footnote{The Higgs mass is maximized for
  $|A_t|/m_{\tilde{t}} \simeq 2$, which corresponds to the maximal-mixing
  scenario.  In this case, $m_{\tilde{t}}$ can be as small as
  $\sim$ 300 GeV without violating the bound on $m_h$ from
  LEP. However it is not trivial to achieve the maximal-mixing
  scenario in models. For more details see {\it e.g.} the discussion in
  Refs.~\cite{Dermisek:2006ey,Dermisek:2007yt}. }

Before we connect the discussion of naturalness of EWSB and the Higgs mass it is worthwhile to note an indirect prediction for the SM Higgs mass 
from precision electroweak data. The latest best fit is obtained for $m_h = 90$ GeV with an experimental uncertainty of +36 and -27 GeV (at 68 percent confidence level)~\cite{lepewwg}. Excluding the most discrepant measurement, the forward-backward asymmetry for $b$ quarks -- $A^b_{FB}$ (with a pull of about 3$\sigma$), the best fit value for the Higgs mass is significantly lower which increases the tension with the direct search limit~\cite{Chanowitz:2002cd}.

\subsection{Fine Tuning Problem}

From the discussion above we can easily see
the tension between the direct search bound on the Higgs mass and naturalness 
of electroweak symmetry breaking in the MSSM.

In order to push the Higgs mass above the
LEP limit, 114.4 GeV, assuming the typical mixing in the stop sector, the stop masses
have to be $\gsim 1\tev$. With 1 TeV stop masses one would expect the mass of the Z boson to be of order 1 TeV, see Eq. (\ref{eq:expectation}). Or alternatively, with the natural expectation for the stop masses from electroweak symmetry breaking, one would obtain the mass of the Higgs boson $\lesssim 100$ GeV, which is ruled out by LEP limit. 

Stop masses of order 1 TeV mean that there is a term of order $(1 \; {\rm TeV})^2$ on the right hand side of Eq. (\ref{eq:MZ_gut}), or that the $\delta m_{H_u}^2$ in Eq. (\ref{eq:MZdel}) is of order $(1 \; {\rm TeV})^2$.\footnote{This result can be obtained by integrating the RG equation for $H_u$ keeping only the term proportional to stop masses, $ \delta m_{H_u}^2 \simeq  - 3 \lambda_t^2 /(4\pi^2 ) \,  m_{\tilde{t}}^2 \, \log (M_{GUT} /M_Z )$.  
Numerically the loop factor times large log is of order one and assuming the stop masses do not change significantly in the RG evolution we find $\delta m_{H_u}^2 \simeq - m_{\tilde{t}}^2$. This gives the correct answer, however it is somewhat oversimplified because it hides (the most important) effect of the gluino mass on the evolution of stop masses and consequently the Higgs soft mass parameter which is clearly visible in Eqs. (\ref{eq:mstop_gut}) and (\ref{eq:MZ_gut}) and can be obtained only by solving coupled RG equations.}  
Then in order to obtain $M_Z^2 = (91 \; {\rm GeV})^2$ on the left-hand side of these equations we need to cancel the large contribution on the right-hand side resulting from the stop masses by something else with about 1\% precision. In Eq.  (\ref{eq:MZdel}) it can be either the boundary condition at the GUT scale or the $\mu$ parameter. In both cases small changes in boundary conditions would generate a very different value for the
EW scale. 

The situation improves when considering large mixing in the stop sector. The mixing is controlled by  
the ratio of $A_t - \mu \cot \beta$ and $m_{\tilde{t}}$. Since we consider parameter space where $\mu$ is 
small to avoid fine tuning and $\tan \beta \gsim 5$ in order to maximize the tree level 
Higgs mass, see Eq. (\ref{eq:mh_mix}), the mixing is simply given by $A_t / m_{\tilde{t}}$.
It was realized that mixing $A_t(M_Z) / m_{\tilde{t}} (M_Z) \simeq \pm 2$ maximizes the Higgs mass for 
given $m_{\tilde{t}}$~\cite{Carena:2000dp}, while still satisfying constraints to avoid charge and color breaking (CCB) minima~\cite{LeMouel:2001ym}. 
Using {\it FeynHiggs-2.2.10} we find that $m_{\tilde{t}} (M_Z) \simeq 300$ GeV and $|A_t (M_Z)| = 450$ 
GeV (for $\tan \beta \gsim 50$),  $|A_t (M_Z)| = 500$ (for any $\tan \beta \gsim 8$) or  $|A_t (M_Z)| = 600$ GeV (for  $\tan \beta$ as small as 6) satisfies the LEP limit on the Higgs 
mass.  Therefore large stop mixing, $|A_t(M_Z) / m_{\tilde{t}} (M_Z)| \gsim 1.5$ is crucial for satisfying the LEP limit  with light stop masses (the physical stop mass in this case can be as small as current experimental bound, $m_{\tilde{t}_1} \gsim 100$ GeV).
 Decreasing the mixing requires increasing of $m_{\tilde{t}}$ and finally we end up with 
 $m_{\tilde{t}} \gsim 1$ TeV for small mixing.
 
 In the limit when the stop mass, $m_{\tilde{t}} (M_Z) \simeq 300$ GeV, originates mainly from $M_3$, from Eq.~(\ref{eq:mstop_gut}) we see we need $M_3 \simeq 130$ GeV.
Then Eq.~({\ref{eq:At_gut}) shows that the necessary $|A_t(M_Z)| \simeq 500$ GeV 
is obtained only when $A_t \lsim -1000$ GeV or $A_t \gsim 4000$ GeV at the GUT scale, 
in both cases it has to be signifficantly larger than other SSBs.
The contribution from the terms in Eq. (\ref{eq:MZ_gut}) containing $M_3$ and $A_t$ is at least $(600 \, {\rm GeV})^2$ and therefore large radiative correction have to be 
cancelled either by $\mu^2$ or $m_{H_u}^2 (M_{GUT})$. 
If $m_{\tilde{t}}$ is not negligible at the GUT scale, $M_3$ can be 
smaller, but in this case we need even larger $A_t$ and the conclusion is basically the same. 
Thus, although the situation improves by considering large $A_t$ term, we still need at least 3 $\%$ fine tuning~\cite{Dermisek:2006ey}.\footnote{For additional discussion of the connection between the stop sector and the Higgs boson mass and related issues of fine tuning of EWSB  see also Refs.~\cite{Essig:2007vq,Essig:2007kh}.}

To summarize, in the MSSM with unrelated soft-susy-breaking terms the LEP limit on the Higgs mass requires cancellations between soft-susy-breaking parameters and/or the $\mu$ parameter at the level of $\sim 1 \%$ (optimizing the mixing in the stop sector reduces the required cancellations to $\sim 3 \%$).

\subsection{Naturalness as a guide}

Naturalness of EWSB in supersymmetric models is the only reason to expect superpartners at the LHC. If naturalness or the fine-tuning criterion are not relevant,  models like MSSM can be realized in nature with superpartners beyond the LHC reach, $\gtrsim 3$ TeV,  while still providing all the virtues discussed in the introduction, namely gauge coupling unification\footnote{Somewhat heavy superpartners typically lead to larger splitting between masses of squarks and sleptons which can even improve the precision with which the gauge couplings unify through the weak scale threshold corrections~\cite{Dermisek:2004mh}.}, radiative EWSB, even the explanation of the observed dark matter density. And all this requires only fine tuning at the level of $0.1 \%$. 

The acceptable level of fine tuning is clearly a subjective criterion\footnote{For an interesting review of various views of naturalness see {\it e.g.} Ref.~\cite{Giudice:2008bi}.} and it is certainly not a reason to abandon some models or ideas. After all, the smallness of the electroweak scale might be just an accident without any explanation, or simply the whole idea of low scale supersymmetry is wrong and there is a different mechanism realized in nature that is responsible for EWSB. 

For the purposes of this review we will nevertheless assume that the low scale supersymmetry is the scenario responsible for EWSB, and more importantly, the EWSB happens in a natural way without necessity of fine-tuning between radiative corrections and boundary values of relevant parameters. This is the premise of this review and  we will focus on the consequences this assumption has for 
the Higgs sector and/or SUSY breaking scenarios. We will see that the naturalness criterion will lead us to consider models in which the Higgs boson decays in an unusual way or models 
with unusual pattern of soft supersymmetry breaking parameters.\footnote{Due to limited space we do not discuss models  that focus on increasing the mass of the Higgs boson by additional interactions, e.g. gauge interaction, see Refs.~\cite{Batra:2003nj,Maloney:2004rc}.}

\section{Models with Non-Standard Higgs Decays}

As we saw the fine-tuning problem in EWSB in the MSSM doesn't result from non-observation of superpartners (current experimental limits on relevant superpartners, see Eq. (\ref{eq:expectation}),  are not far from the EW scale) but rather from the non-observation of the Higgs boson. This suggests that the fine-tuning problem could be eliminated if one can avoid the LEP limits on the Higgs mass~\cite{Dermisek:2005ar}.

The solution to the fine-tuning problem in models in which the
SM-like Higgs decays dominantly in a different way is straightforward.
If the $h \to b \bar b$ decay mode is not dominant the Higgs boson
does not need to be heavier than 114 GeV, it can be as light as the
typical Higgs mass or even ligter depending on the limits placed on
the dominant decay mode. If this limit is $\leq 100\gev$, there is no
need for large superpartner masses and superpartners can be as light
as current experimental limits allow~\cite{Dermisek:2005ar}. For a review of the
experimental limits on the mass of the SM-like Higgs boson in various
decay modes see Ref.~\cite{Chang:2008cw}. Quite surprisingly, only if the Higgs decays primarily to
two or four bottom quarks, two jets, two taus or to an invisible
channel (such as two stable LSP's), is the LEP limit on $m_h$ above
$100\gev$. LEP limits on $m_h$ for all other decay modes are below $90\gev$
and would therefore not place a constraint on
superpartner masses.  Thus, models where these alternate (less constrained) decay modes
are dominant provide a
solution to the fine tuning problem.\footnote{In specific models,
  avoiding the fine tuning problem might require another tuning of
  parameters in order to make an alternate decay mode for the Higgs
  boson dominant, see e.g. Refs.~\cite{Schuster:2005py,Dermisek:2006wr} for the discussion of these issues in the NMSSM.}

\subsection{NMSSM with a  light  singlet-like CP odd Higgs}

The situation we have just discussed already happens in the simplest extension of the MSSM - 
the next-to-minimal supersymmetric model (NMSSM). The NMSSM  introduces an additional singlet super-field with the following couplings
\beq
W \supset \lambda \widehat{S}
\widehat{H}_u \cdot \widehat{H}_d + \frac{1}{3} \kappa \widehat{S}^3 \ .
\eeq 
The vacuum expectation value of the singlet effectively generates the $\mu$ term.
The Higgs spectrum consist of three CP even Higgses, $h_{1,2,3}$, two CP odd Higgses, $a_{1,2}$
and a pair of charged Higgs bosons.
It was found that in the NMSSM the SM-like Higgs boson
can be light, $m_{h_1} \sim 100$ GeV, as is predicted from natural EWSB with a generic SUSY
spectrum. The Higgs in this scenario
 decays partly via
$h_1 \to b\bar b$  but dominantly into two CP-odd
Higgs bosons, $h_1 \to a_1 a_1$, where $m_{a_1}<2m_b$ so that $a_1 \to \tau^+ \tau^-$
(or light quarks and gluons) decays are dominant~\cite{Dermisek:2005ar,Dermisek:2007yt}.
Further, since the $ZZh$ and $WWh$ couplings are
very SM-like, these scenarios give excellent agreement with precision
electroweak data.

Besides alleviating or completely removing the fine tuning problem the
possibility of modified Higgs decays is independently supported
experimentally. The largest excess ($2.3 \sigma$) of Higgs-like events
at LEP is that in the $b\bar b$ final state for reconstructed
$M_{b\bar b}\sim 98 \gev$.  The number of excess events is roughly
10\% of the number of events expected from the standard model with a
98 GeV Higgs boson.  Thus, this excess cannot be interpreted as the
Higgs of the standard model or the SM-like Higgs of the
MSSM.\footnote{In the MSSM this excess can be explained by the Higgs
  with highly reduced coupling to $ZZ$. This explanation doesn't
  remove the fine tuning problem since it is the heavy Higgs which is
  SM-like and has to satisfy the 114 GeV limit. For a detailed
  discussion and references, see Ref.~\cite{Dermisek:2007ah}.}
However, this excess is a perfect match to the idea of non-standard
Higgs decays.  As we have discussed, from natural EWSB we expect the
SM-like $h$ to have mass very near 100 GeV, and this is possible in
any model where the SM-like Higgs boson decays mainly in a mode for
which the LEP limits on $m_h$ are below $90\gev$.  The $h \to b \bar
b$ decay mode will still be present, but with reduced branching ratio.
Any $\br(h \to b \bar b) \lsim 30\%$ is consistent with experimental
limits for $m_h \sim 100$ GeV. Further, $\br(h \to b \bar b) \sim
10\%$ with $m_h\sim 100\gev$ provides a perfect explanation of the
excess.  This interpretation of the excess was first made in the NMSSM scenario discussed above 
with the $h_1 \to a_1a_1 \to \tau^+ \tau^- \tau^+ \tau^- $ mode being
dominant~\cite{Dermisek:2005gg}, but it clearly applies to a wide
variety of models. The NMSSM scenarios with minimal fine-tuning, and thus with the SM-like Higgs mass at $\sim 100$ GeV, nicely explain the excess, see Fig.~\ref{zbblimits} (from
Ref.~\cite{Dermisek:2007yt}). 

In summary, models with non-standard Higgs decays can avoid the fine-tuning problem by allowing the Higgs boson mass to be  $\sim
100\gev$, value predicted from natural EWSB, while at the same time the
subdominant decay mode, $h \to b \bar b$, with $\sim 10\%$ branching
ratio can explain the largest excess of Higgs-like events at LEP at
$M_{b\bar b}\sim 98$ GeV. A SM-like $h$ with $m_h\sim 100\gev$ is also
nicely consistent with precision electroweak data. Besides the NMSSM scenario discussed above, non-standard Higgs decays occur  in a variety of models, for a review see Ref.~\cite{Chang:2008cw}. 
An interesting model featuring all the virtues of the above scenario with  the SM-like Higgs decaying into four gluons appeared recently~\cite{Bellazzini:2009xt}.

\subsection{Light  doublet-like  CP odd Higgs at small $\tan \beta$}

An interesting variation of the above NMSSM scenario is the scenario with a {\it doublet-like} CP odd Higgs bellow the $b \bar b$ threshold.
For small  $\tan \beta$, $\tan \beta \lesssim 2.5$, it is  the least constrained (and only  marginally ruled out) in the MSSM, and thus easily viable in simple extensions of the MSSM~\cite{Dermisek:2008id}. Surprisingly the  prediction from this region is that
all the Higgses resulting from two Higgs doublets: $h$, $H$, $A$ and $H^\pm$ could have been produced already at LEP or the Tevatron, but would have escaped detection because they decay in modes that have not been searched for or the experiments are not sensitive to.

The heavy CP even and the CP odd Higgses could have been produced at LEP in $e^+ e^- \to H A$ but they would avoid detection because $H$ dominantly decays to $ZA$ - the mode  that has not been searched for. 
The charged Higgs is also very little constrained and
up to $\sim 40 \%$  of top quarks produced at the Tevatron could have decayed into charged Higgs and the $b$ quark since the dominant decay mode for the charged Higgs $H^\pm \to W^{\pm \star} A$ with $A \to c \bar c$ or $\tau^+ \tau^-$ was not searched for either. In addition the charged Higgs with the properties emerging in this scenario and mass close to the mass of the $W$ boson
could explain the $2.8 \sigma$ deviation from lepton universality in $W$ decays measured at LEP~\cite{:2004qh} as discussed in Ref.~\cite{Dermisek:2008dq}. 

The mass of the light CP even Higgs is the only problematic part in this scenario. In the MSSM in this region of the parameter space we find  $m_h \simeq 40 - 60 $ GeV,  and thus $h \to AA$ decay mode is open and generically dominant. However the decay mode independent limit requires the SM like Higgs to be above 82 GeV which rules this scenario out in the MSSM,
since $m_h$ cannot be pushed above 82 GeV by radiative corrections.
There are however various ways to increase the mass of the SM-like Higgs boson in extensions of the MSSM. A simple
possibility is to consider singlet extensions of the MSSM containing $\lambda S H_u H_d$ term in the superpotential.
It is known that this term itself contributes $\lambda^2 v^2 sin^2 2 \beta$, where v = 174 GeV, to the mass squared of
the CP even Higgs~\cite{Ellis:1988er} and thus can easily push the Higgs mass above the decay-mode independent limit, 
$82$ GeV. Note, this contribution
is maximized for $\tan \beta \simeq 1$.\footnote{Sometimes it is argued that the extra contribution to the Higgs mass in the NMSSM makes the NMSSM less fine-tuned than the MSSM since the LEP limit 114 GeV is easier to satisfy. This effect is not very significant however. For large $\tan \beta$ the extra contribution is negligible and for small $\tan \beta$ we are loosing the tree level contribution we have in the MSSM, see Eq. (\ref{eq:mh_mix}). Optimizing  $\tan \beta$ improves the naturalness of EWSB in the NMSSM compared to the MSSM only marginally~\cite{Dermisek:2007yt}.} Thus it is not surprising that the scenario with a light doublet-like CP-odd Higgs boson is phenomenologically viable in the NMSSM~\cite{Dermisek:2008uu} but clearly since the light CP-odd Higgs comes from the two Higgs doublets it is not limited to singlet extensions of the MSSM.

\subsection{Higgs as a Link to  New Sectors}

The extra singlet in the NMSSM 
does not spoil any of the virtues of the MSSM, including the
possibility of gauge coupling unification and matter particles fitting
into complete GUT multiplets. 
It is also very interesting to consider a possibility that besides the extra singlet there is a whole extra ($E$) sector of 
particles (for simplicity we assume that these particles are scalars) that are singlets under standard model gauge symmetry and couple to the MSSM sector only through the Higgs fields.
Such couplings would have a negligible effect on the
phenomenology involving SM matter particles, whereas they can
dramatically alter Higgs physics. For example, they would allow the
lightest CP-even Higgs boson $h$ to decay into two of the $E$-particles if the $E$-particles
are light enough.

If the $h$ decays to two lighter $E$-particles, then the strategy for
Higgs discovery will depend on the way the $E$-particles appearing in
the decays of the $h$ themselves decay. They might decay predominantly
into other stable $E$-particles, in which case the MSSM-like $h$
decays mainly invisibly. More typically, however, the $E$-particles acquire a tiny coupling to SM particles 
via interactions with the MSSM Higgs fields (scalar mass eigenstates are generically mixed).
In this case, light $E$-particles 
will decay into $b \bar b$, $\tau^+ \tau^-$ or other quarks or
leptons depending on the model. 
Although $E$-particles would have
small direct production cross sections and it would be difficult to
detect them directly, their presence would be manifest through the
dominant Higgs decay modes being $h \to 4f$, where $4f$ symbolically
means four SM particles, \eg\ $b \bar b b \bar b$, $b\bar
b\tau^+\tau^-$, $\tau^+ \tau^- \tau^+ \tau^-$, $4 \gamma$ and so on.
The situation can be even more complicated if the $h$ decays to
$E$-particles that themselves decay into other $E$-particles, $\dots$,  which in
turn finally decay into SM particles. In this case the SM-like Higgs boson would cascade decay in the $E$-sector till it decays to the lightest $E$-particles, which finally decay to SM particles. Depending on the mass of the lightest $E$-particle this would look like the SM-like Higgs is decaying into a large number of light jets, or a large number of muons, or a large number of electrons, or in case the  lightest $E$-particle is below  $2 m_e$ threshold the only way it can decay is into two photons, so effectively the SM-like Higgs boson would burst into a large number of soft photons~\cite{superB-talk}. These events would be quite spectacular, e.g. the decay of the Higgs boson to many photons would light up a large portion of the electromagnetic calorimeter.
Such signatures are not very usual in particle physics which is probably the reason 
these were never looked for. 

The light CP-odd Higgs of the NMSSM or other extended models might be within the reach of current B factories where it can be produced  
in Upsilon decays, $\Upsilon \to A \gamma$~\cite{Dermisek:2006py}. The same applies to light $E$-sector particles\footnote{The light CP-odd Higgs of the NMSSM is an example of an $E$-sector with only one particle.} with all the possible spectacular decay modes mentioned above~\cite{superB-talk}.

\section{Unusual Supersymmetry}

Another possibility to reconcile non-observation of the Higgs boson with naturalness of electroweak symmetry breaking assumes that
the MSSM is the correct description of nature at the TeV scale but the usual assumptions we make 
about SUSY breaking are not correct.  In this case in order to radiatively trigger EWSB and satisfy the limit on the Higgs mass 
SUSY has to be quite special.

Although $M_Z$ results from cancellations between SSB parameters in this case, 
it does not necessarily mean that the Z mass is  fine-tuned. 
SUSY breaking scenarios typically  produce SSBs 
which are related to each other in a specific way and so they
should not be treated as independent parameters.
In such a case, the discussion around Eq. (\ref{eq:MZ_gut}) is not relevant to assess the level of fine tuning, however it still 
can be used to explore
 relations between SSBs that have to be generated, 
should the $M_Z$ emerge in a natural way.

A simple example of a relation between SSBs that allows relevant SSBs to be large without  significantly contributing to $M_Z$ is $m_{\tilde{t}} = m_{H_u}$.
In this case the contribution from $m_{\tilde{t}}$ approximately  cancels the contribution from  $m_{H_u}$ in Eq. (\ref{eq:MZ_gut}) and this cancelation is more precise for large $\tan \beta$ as discussed in Ref.~\cite{Feng:1999mn}. Thus scalars can be much heavier than gauginos without contributing more to $M_Z$.

Even larger hierarchy between scalars and gauginos can be achieve with more optimized relations between SSBs: $A_0 \simeq -2 m_{16}$ and $ m_{10} \simeq \sqrt{2}  m_{10}$ with $M_{1/2} \ll m_{16}$, where $A_0$ is the universal soft trilinear coupling at the GUT scale, and  $m_{16}$ and $ m_{10}$ are universal masses of superpartners of three families and the Higgs masses respectively. These conditions, motivated by SO(10) grand unified theories, were found to radiatively generate the largest hierarchy between the first two generations of scalars and the third one~\cite{Bagger:1999sy} and were also considered for having attractive features, including the possibility of Yukawa coupling unification~\cite{Blazek:2001sb,Blazek:2002ta} and maximizing the suppression of the proton decay~\cite{Dermisek:2000hr}. However it should be stressed that these type of relations could reduce fine-tuning of EWSB only if they can be realized as a unique (or at least a discrete) possibility in a given SUSY breaking scenario, which proved not to be easy.

More recent example is a proper mixture of  anomaly and modulus mediation
~\cite{Choi:2005uz,Choi:2005hd,Kitano:2005wc,Lebedev:2005ge} which produces boundary conditions such that the initial value of $m_{H_u}^2$ is canceling most of the contribution from running.

All the above mentioned examples have one thing in common: the boundary conditions for soft SUSY breaking parameters generated by some mechanism are such that the boundary condition for $m_{H_u}^2$ is canceling the accumulated radiative correction $\delta m_{H_u}^2$ from the RG evolution, see Eq. (\ref{eq:MZdel}). Clearly such solutions must exist, and some of them might be even realized in simple models but if this cancellation is required to be very precise then there is a  ``coincidence" problem: the relations that have to be satisfied between SSBs in order to recover the correct $M_Z$ depend on the energy interval SSBs are going to be evolved over. Therefore a SUSY breaking scenario would have to  know that SSBs will evolve according to MSSM RG equations, and exactly from e.g. $M_{GUT}$ to $M_Z$. For example, the top Yukawa coupling is responsible for driving the $ m_{H_u}^2$ to negative values and  $\delta m_{H_u}^2$  directly depends on its value. In models we envisioned so far the generated boundary condition for $ m_{H_u}^2$ does not depend on the value of the top Yukawa coupling and thus it would be a ``coincidence" that after RG evolution $\delta m_{H_u}^2$ precisely cancels the boundary condition at a high scale. In other words the $M_Z$ in this case is highly sensitive to the value of the top Yukawa coupling.\footnote{Similar discussion applies to the sensitivity of $M_Z$  to $\alpha_3$.}

In this section we will discuss a possibility that $\delta m_{H_u}^2 \simeq O(M_Z)$. Therefore none of the terms in Eq. (\ref{eq:MZdel}) are large and no cancellation, or fine-tuning, is necessary to reproduce the correct value of $M_Z$. The cancellations (a relations between SSBs supplied by a given model) might still be required to achieve  $\delta m_{H_u}^2 \simeq O(M_Z)$. Nevertheless these scenarios could avoid the coincidence problem mentioned above.
A typical feature of these scenarios is the reduced sensitivity of $M_Z$ to dimensionless couplings, especially the top Yukawa coupling. Additional interesting feature is that  a large hierarchy between the fundamental scale at which SSB are generated, e.g. the GUT scale or the Planck scale, and the EW scale is required in order for these scenarios to be viable.

\subsection{Negative stop masses squared}

A simple way to achieve  $\delta m_{H_u}^2 \simeq O(M_Z)$ is to consider negative stop masses squared at a high scale~\cite{Dermisek:2006ey}.
If we allow negative stop masses squared at the GUT scale several interesting things happen simultaneously. First of all, from Eq.~(\ref{eq:mstop_gut}) we see that unless $m_{\tilde{t}}$ is too large compared to $M_3$ it will run to positive values at the EW scale. At the same time the contribution to $m_{H_u}^2$ from the energy interval where $m_{\tilde{t}}^2 <0 $ partially or even exactly cancels the contribution from the energy interval where $m_{\tilde{t}}^2 >0 $ and so the EW scale value of $m_{H_u}^2$ can be arbitrarily close to the starting value at $M_{GUT}$, see Fig.~\ref{fig:RGrunning}. 
From Eq.~(\ref{eq:MZ_gut}) we see that this happens for $m_{\tilde{t}}^2 \simeq - 4 M_3^2$ (neglecting $A_t$).
No cancellation between initial value of $m_{H_u}^2$ (or $\mu$) and the contribution from the running is required.
And finally, from Eqs.~(\ref{eq:mstop_gut}) and  (\ref{eq:At_gut}) we see that the stop mixing is typically much larger than in the case with positive stop masses squared. For positive (negative) stop masses squared we find  $|A_t(M_Z) / m_{\tilde{t}} (M_Z)| \lsim 1$  $(\gsim 1)$
starting with $A_t = 0$ and small $ m_{\tilde{t}}$ at the GUT scale. Starting with larger $ m_{\tilde{t}}$ the mixing is even smaller (larger) in the positive (negative) case. Therefore large stop mixing at the EW scale is generic in this scenario and actually it would require very large GUT scale values of $A_t $ to end up with small mixing at the EW scale.
\begin{figure}
\includegraphics[width=2.45in]{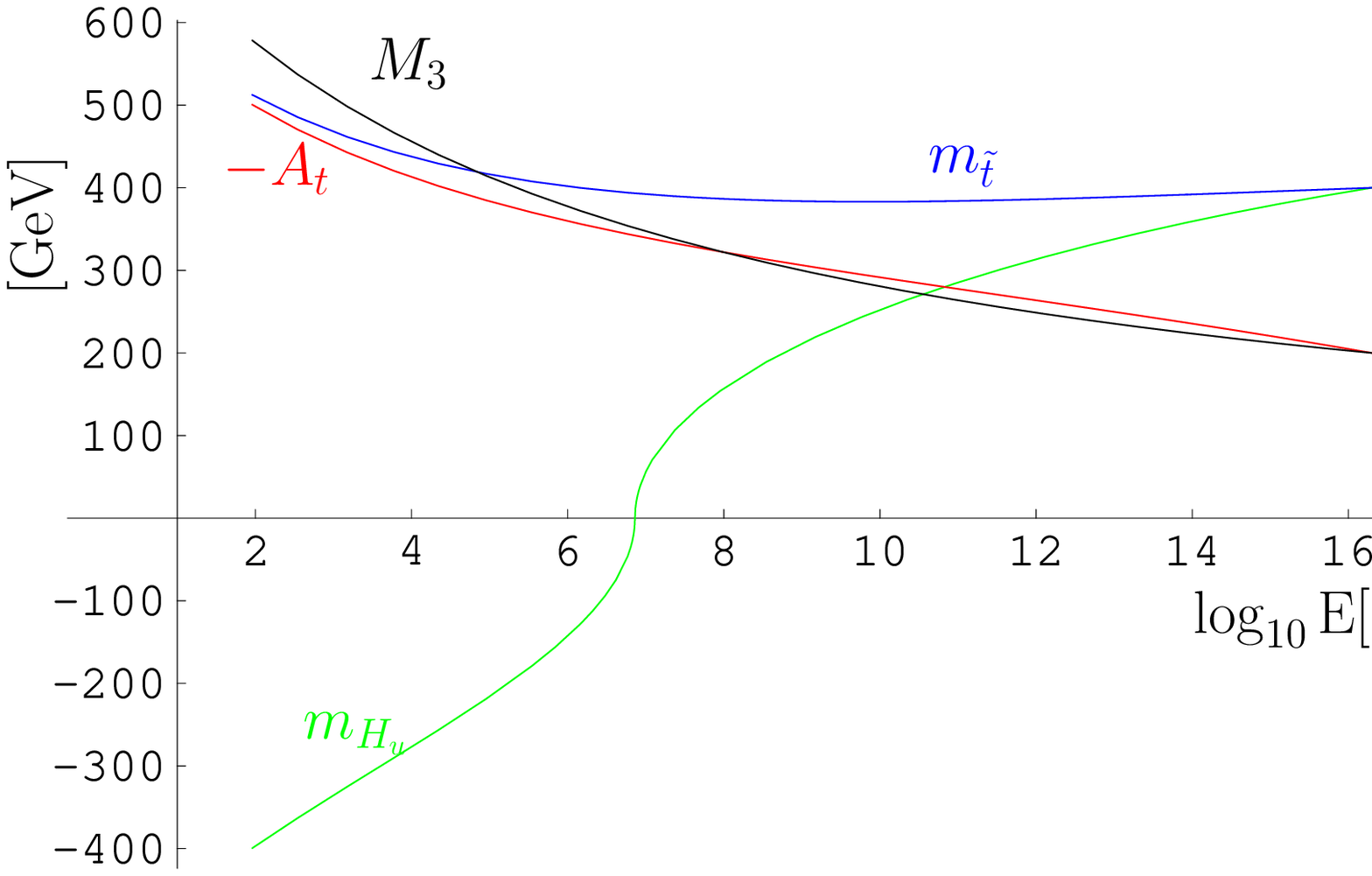}
\includegraphics[width=2.45in]{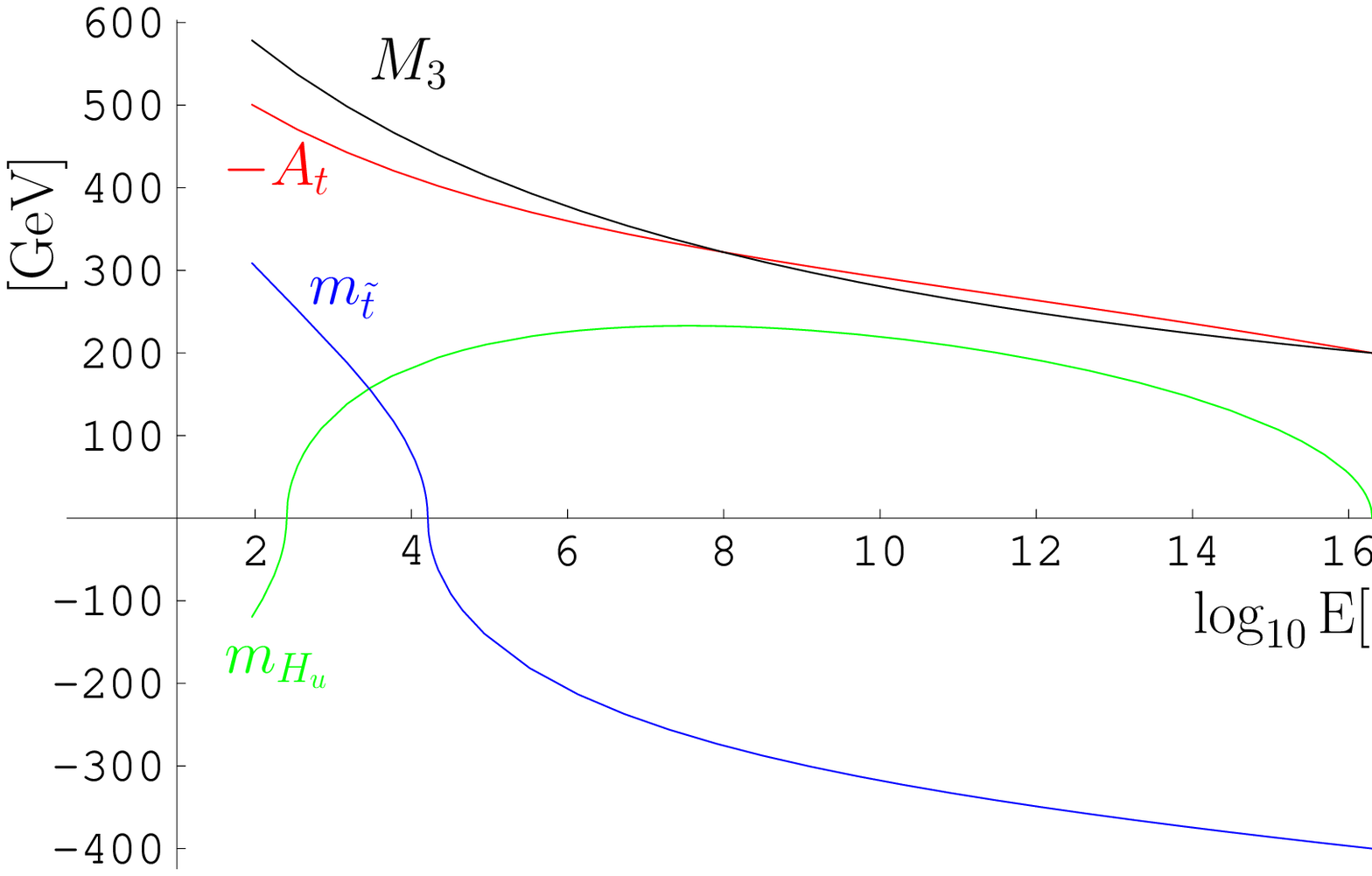}
\caption{Right: renormalization group running of relevant SSBs for $\tan \beta = 10$ and GUT scale boundary conditions: $-A_t = M_3 = 200$ GeV, $m_{\tilde{t}}^2 = -(400 \, {\rm GeV})^2$ and $m_{H_u}^2 = 0 \, {\rm GeV}^2$. Left: the same for $m_{\tilde{t}}^2 = + (400 \, {\rm GeV})^2$; the $m_{H_u}^2$
is chosen to be the same as   $m_{\tilde{t}}^2$ in this case in order to keep the same y-axis scale in both plots. 
In order to have both mass dimension one and two parameters on the same plot and keep information about signs, we define $m_{H_u} \equiv m_{H_u}^2/\sqrt{|m_{H_u}^2|} $ and $m_{\tilde{t}}\equiv m_{\tilde{t}}^2/\sqrt{|m_{\tilde{t}}^2|} $. }
\label{fig:RGrunning}
\end{figure}

It turns out that in the region where $m_{H_u}^2$ gets negligible contribution from running, the radiatively generated stop mixing is close to maximal even when starting with negligible mixing at the GUT scale. In this case, comparing Eqs.~(\ref{eq:mstop_gut}) and (\ref{eq:At_gut}), we find~\footnote{To be more precise the generated mixing is somewhat larger than that shown in this equation, since we should minimize the potential at the SUSY scale $\sim m_{\tilde{t}}$ and should not run SSBs all the way to $M_Z$ (see~Fig.~\ref{fig:RGrunning}).}
\begin{equation}
A_t(M_Z) / m_{\tilde{t}} (M_Z) \simeq -1.5 + 0.2 A_t/M_3.
\label{eq:generatedAt}
\end{equation}
Slightly more negative stop masses squared at the GUT scale would result in maximal stop mixing at the EW scale even when starting with negligible $A_t$. Nevertheless
the example in~Fig.~\ref{fig:RGrunning} with simple GUT scale boundary conditions  already
leads to EW scale parameters
$m_{\tilde{t}}(M_Z) \simeq 300$ GeV and $A_t(M_Z) = -500$ GeV producing sufficiently heavy  Higgs boson, $m_h \simeq 115.4$ GeV. Small variations of GUT scale parameters, including positive or negative values of $m_{H_u}^2$,  would produce similar results and scaling all parameters up would lead to larger Higgs mass.
 
In a theory which predicts $m_{\tilde{t}}^2 \simeq - 4 M_3^2$, the
fine tuning problem is entirely solved. The contribution to $m_{H_u}^2$ from the running is negligible and the ${\cal O}(M_Z^2)$ values of $m_{H_u}^2$ and $\mu^2$ at the GUT scale naturally result in the correct $M_Z$. However, the absence of fine tuning is quite robust and the relation above does not have to be satisfied very precisely.  
Requiring 
fine tuning less than $10 \%$ the stop masses squared have to be generated between  $\simeq - 3  M_3^2$ and $\simeq - 5  M_3^2$ for $M_3 \simeq 200$ GeV.
This interval is shrinking with increasing $M_3$ which is a sign of the coincidence problem discussed above.

Models with negative stop masses at a high scale also reduce the sensitivity of the EW scale to the top Yukawa coupling. This can be qualitatively understood from the fact that it is the top Yukawa coupling that drives the $m_{H_u}^2$ to positive values in the energy interval where stop masses are negative, and it is the same top Yukawa coupling that drives it to negative values in the energy interval where stop masses are positive. Thus larger top Yukawa coupling would drive the  $m_{H_u}^2$ to higher positive values but then, when stop masses turn positive,  it would drive it to negative values faster.

Negative  stop masses at a high scale can be realized e.g. in models with  gauge messengers~\cite{Dermisek:2006qj}.

\subsubsection{Gauge Messenger Model}

In the 
simple gauge messenger  model, based on SU(5) supersymmetric GUT with a minimal particle content, it is assumed that an adjoint chiral superfield, $\hat \Sigma$, gets a vacuum 
expectation value in both its scalar and auxiliary components: $\langle \hat \Sigma \rangle = (\Sigma + \theta^2 F_\Sigma ) \times \rm{diag} (2,2,2,-3,-3)$. The vev in the scalar component, $\Sigma \simeq M_G$, gives supersymmetric masses to  X and Y gauge bosons and gauginos and thus breaks SU(5) down to the standard model gauge symmetry. The vev in the F component, $F_\Sigma$, splits masses of heavy gauge bosons and gauginos and breaks suppersymmetry. The SUSY breaking is communicated to MSSM scalars and gauginos through loops involving these heavy gauge bosons and gauginos which play the role of messengers (the messenger scale is the GUT scale). The gauge messenger model is very economical, all gaugino and scalar masses are given by one parameter, 
\bea
M_{\rm SUSY} & = & \frac{\alpha_G}{4\pi} \f{|F_\Sigma|}{M_G},
\eea
and it is phenomenologically viable~\cite{Dermisek:2006qj}.
The characteristic features are:
negative and non-universal squark and slepton masses squared at the
unification scale, non-universal gaugino masses, and sizable soft-trilinear
couplings. The combination of a large negative top soft trilinear coupling and negative stop masses square lead 
to close to maximal
mixing scenario for the Higgs mass and reduce the fine-tuning
of electroweak symmetry breaking. 
In this scenario,
all soft supersymmetry breaking parameters at the unification scale can  be smaller than 400 GeV and
all the superpartners can be lighter than 400 GeV and still satisfy
all the limits
from direct searches for superpartners and also the limit on the Higgs mass.

For soft SUSY breaking parameters in extended models see Ref.~\cite{Dermisek:2006qj} and the discussion of dark matter can be found in Ref.~\cite{Bae:2007pa}.

\subsection{Hypercharge Mediation of SUSY Breaking}

Another scenario with  $\delta m_{H_u}^2 \simeq O(M_Z)$ is the hypercharge mediation of SUSY breaking~\cite{Dermisek:2007qi}. 
 It is a string motivated scenario which uses a similar setup envisioned for the anomaly mediation~\cite{Randall:1998uk,Giudice:1998xp} and it is also related to the Z'-mediation~\cite{Langacker:2007ac}.

 The scenario is characterized by the bino mass $M_1$ being the only soft-susy breaking term generated at a high-scale.
 In the RG evolution to the weak scale,
all scalar masses (including the Higgs masses) receive a contribution from the bino mass. This positive contribution dominates at the beginning of the RG evolution. 
Once sizable scalar masses are developed, 
the negative contribution from Yukawa couplings 
becomes important and can overcome the contribution from the bino mass. 
In pure hypercharge mediation, the left-handed stop mass 
squared would be driven to negative values, because out of all scalars 
its hypercharge is the smallest
and its Yukawa coupling is the largest. All other squarks and sleptons 
remain positive, see Fig~\ref{fig:RG_HuQL} (from Ref.~\cite{Dermisek:2007qi}).\footnote{Depending on $\tan \beta$ and the scale at which the SUSY breaking is
communicated to the MSSM sector also the up-type Higgs mass squared,
$m_{H_u}^2$ can be driven to negative values. However, unless the
starting scale is very large (above the GUT scale) $m_{H_u}^2$
remains small and positive at the weak scale.} 
The wino and gluino masses receive a contribution from the bino mass at 
the two loop level. 

\begin{figure}
\centerline{\includegraphics[width=3.4in]{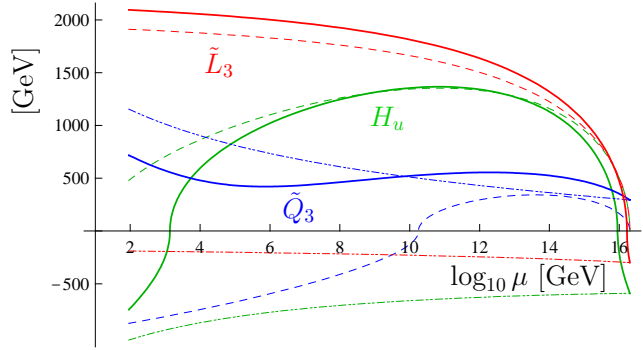}}
\caption{Renormalization group running of $m_{H_u}$ (green), $m_{Q_3}$ (blue) and $m_{L_3}$ (red)
 for $\tan \beta = 10$,
$ m_{3/2} = 50$ TeV and $\alpha = 0.2$ for $M_\star = M_{GUT}$.
We define $m_{H_u} \equiv m_{H_u}^2/\sqrt{|m_{H_u}^2|} $ and similarly for $m_{Q_3}$ and $m_{L_3}$. 
The contribution of pure hypercharge mediation 
is given by dashed lines and  
the separate contribution from anomaly mediation 
is represented by the corresponding dotted lines.
}
\label{fig:RG_HuQL}
\end{figure}

Due to negative stop masses at the EW scale
the scenario is not viable but  small contribution to gaugino masses $O(100 \, {\rm GeV})$ from 
other sources can easily cure this problem.  
For example one can consider a subleading contribution from anomaly mediation~\cite{Dermisek:2007qi}, see Fig~\ref{fig:RG_HuQL} (from Ref.~\cite{Dermisek:2007qi}). Or, one can consider small contribution to gaugino masses from gravity mediation. The gluino mass large enough to satisfy experimental limits is also sufficient to push stops above the experimental limit and drive the $m_{H_u}^2$ to negative values.

When discussing Eq. (\ref{eq:MZ_gut}) we mentioned that the  bino mass enters this formula with a negligible coefficient and we didn't bother to write it down. However, in the hypercharge mediation it is the only nonzero term  (up to possible contribution from other sources). The value of this term  is $- 0.006 \, M_1^2$ and thus $M_1$ of order 4 TeV would not result in fine-tuning larger than 10\%. Thus if the pure hypercharge mediation was a phenomenologically viable model, it would be an example of a model in which there is no cancellation required in EWSB. 
Considering contribution from {\it e.g.} gravity mediation to the other two gaugino masses of order $200 \, {\rm GeV}$ makes the model viable and does not introduce fine tuning larger than $\sim 4 \%$.\footnote{
Heavier bino compared to other gauginos proved to be advantageous in earlier studies of non-universal gaugino mediation~\cite{Baer:2002by,Balazs:2003mm}. However there was no theoretical motivation for such a  choice.} 

\section{Interesting Coincidences?}

The LHC will start running soon and we will finally learn about the mechanism responsible for electroweak symmetry breaking. Based on limited available experimental evidence we have envisioned elegant models that are so ambitious that they could describe Nature even at energies close to the Planck scale. These ideas are based on hints or {\it interesting coincidences} in values of parameters we observe. The most influential such a coincidence is the fact that gauge couplings unify in the MSSM. 
In addition, the Higgs mass squared parameter, $m_{H_u}^2$, is typically the only one driven to negative values at the EW scale and so SUSY breaking generically results in electroweak symmetry breaking.
These can be just coincidences or Nature gave us a hint of what we will really find at the LHC.

Based on similar hints (which are not so clear however) we are guessing how the Higgs sector should look like and how supersymmetry should be broken.
It is certainly at least an interesting coincidence that the natural value of the Higgs mass in the MSSM and simple extensions coincides with the largest excess of Higgs like events at LEP. It can be just a coincidence or maybe it is a hint that the Higgs really is at $\sim 100$ GeV.   

Another coincidence is that in models with negative stop masses squared at a high scale the least fine tuned scenarios automatically maximize the Higgs mass through the mixing in the stop sector. In addition, these models, or in general models with  $\delta m_{H_u}^2 \simeq O(M_Z)$, including the hypercharge mediation, require a large hierarchy (more than 10 orders of magnitude) between the scale at which SUSY is broken (GUT scale or Planck scale) and the EW scale in order to trigger EWSB. 

All these could be just coincidences and some of them certainly are. We put a lot of effort to build a beautiful picture out of what we know. If  none of this is realized in Nature, and we were completely misled, 
 let us hope that Nature prepared for us something even more elegant.

\section*{Acknowledgments}

I would like to thank J.F. Gunion, H.D. Kim and S. Raby for collaboration and countless discussions on the subject presented in this review.


\begin{thebibliography}{0}



\bibitem{Dimopoulos:1981yj}
  S.~Dimopoulos, S.~Raby and F.~Wilczek,
  Phys.\ Rev.\  D {\bf 24}, 1681 (1981).
  
\bibitem{Dimopoulos:1981zb}
  S.~Dimopoulos and H.~Georgi,
  Nucl.\ Phys.\  B {\bf 193}, 150 (1981).
  
\bibitem{Ibanez:1981yh}
  L.~E.~Ibanez and G.~G.~Ross,
  Phys.\ Lett.\  B {\bf 105}, 439 (1981).
  
\bibitem{Sakai:1981gr}
  N.~Sakai,
  Z.\ Phys.\  C {\bf 11}, 153 (1981).
  

\bibitem{Witten:1981nf}
  E.~Witten,
  Nucl.\ Phys.\  B {\bf 188}, 513 (1981).
  
\bibitem{Ibanez:1982fr}
  L.~E.~Ibanez and G.~G.~Ross,
  Phys.\ Lett.\  B {\bf 110}, 215 (1982).
  
\bibitem{Barate:2003sz}
  R.~Barate {\it et al.}  [LEP Working Group for Higgs boson searches],
  Phys.\ Lett.\ B {\bf 565}, 61 (2003)
  [arXiv:hep-ex/0306033].
  

\bibitem{Giudice:2007qj}
  G.~F.~Giudice,
  J.\ Phys.\ Conf.\ Ser.\  {\bf 110}, 012014 (2008)
  [arXiv:0710.3294 [hep-ph]].




\bibitem{Ibanez:1983di}
  L.~E.~Ibanez and C.~Lopez,
  Nucl.\ Phys.\ B {\bf 233}, 511 (1984).

\bibitem{Carena:1996km}
  M.~Carena, P.~Chankowski, M.~Olechowski, S.~Pokorski and C.~E.~M.~Wagner,
  Nucl.\ Phys.\ B {\bf 491}, 103 (1997).
   
  
\bibitem{Okada:1990vk}
  Y.~Okada, M.~Yamaguchi and T.~Yanagida,
  Prog.\ Theor.\ Phys.\  {\bf 85}, 1 (1991).
  
\bibitem{Haber:1990aw}
  H.~E.~Haber and R.~Hempfling,
  Phys.\ Rev.\ Lett.\  {\bf 66}, 1815 (1991).
  
\bibitem{Ellis:1990nz}
  J.~R.~Ellis, G.~Ridolfi and F.~Zwirner,
  Phys.\ Lett.\  B {\bf 257}, 83 (1991).
  
\bibitem{Ellis:1991zd}
  J.~R.~Ellis, G.~Ridolfi and F.~Zwirner,
  Phys.\ Lett.\  B {\bf 262}, 477 (1991).



\bibitem{Dermisek:2007yt}
  R.~Dermisek and J.~F.~Gunion,
  Phys.\ Rev.\  D {\bf 76}, 095006 (2007)
  [arXiv:0705.4387 [hep-ph]].
  
  
\bibitem{Dermisek:2007ah}
  R.~Dermisek and J.~F.~Gunion,
  arXiv:0709.2269 [hep-ph].
  




\bibitem{Heinemeyer:1998yj}
  S.~Heinemeyer, W.~Hollik and G.~Weiglein,
  Comput.\ Phys.\ Commun.\  {\bf 124}, 76 (2000).

\bibitem{Heinemeyer:1998np}
  S.~Heinemeyer, W.~Hollik and G.~Weiglein,
  Eur.\ Phys.\ J.\ C {\bf 9}, 343 (1999).





\bibitem{Dermisek:2006ey}
  R.~Dermisek and H.~D.~Kim,
  Phys.\ Rev.\ Lett.\  {\bf 96}, 211803 (2006)
  [arXiv:hep-ph/0601036].
  


\bibitem{lepewwg}
LEP-EWWG, http://www.cern.ch/LEPEWWG.
 
\bibitem{Chanowitz:2002cd}
  M.~S.~Chanowitz,
  Phys.\ Rev.\  D {\bf 66}, 073002 (2002)
  [arXiv:hep-ph/0207123].


\bibitem{Carena:2000dp}
  M.~Carena et al.,
  Nucl.\ Phys.\ B {\bf 580}, 29 (2000),
and references therein.

\bibitem{LeMouel:2001ym}
C.~Le Mouel,
Phys.\ Rev.\ D {\bf 64}, 075009 (2001).



\bibitem{Essig:2007vq}
  R.~Essig,
  Phys.\ Rev.\  D {\bf 75}, 095005 (2007)
  [arXiv:hep-ph/0702104].
  
\bibitem{Essig:2007kh}
  R.~Essig and J.~F.~Fortin,
  JHEP {\bf 0804}, 073 (2008)
  [arXiv:0709.0980 [hep-ph]].
  



\bibitem{Dermisek:2004mh}
  R.~Dermisek,
  arXiv:hep-ph/0401109.
  
\bibitem{Giudice:2008bi}
  G.~F.~Giudice,
  arXiv:0801.2562 [hep-ph].

\bibitem{Batra:2003nj}
  P.~Batra, A.~Delgado, D.~E.~Kaplan and T.~M.~P.~Tait,
  JHEP {\bf 0402}, 043 (2004)
  [arXiv:hep-ph/0309149].

\bibitem{Maloney:2004rc}
  A.~Maloney, A.~Pierce and J.~G.~Wacker,
  JHEP {\bf 0606}, 034 (2006)
  [arXiv:hep-ph/0409127].
  

\bibitem{Dermisek:2005ar}
  R.~Dermisek and J.~F.~Gunion,
  Phys.\ Rev.\ Lett.\  {\bf 95}, 041801 (2005)
  [arXiv:hep-ph/0502105].


\bibitem{Chang:2008cw}
  S.~Chang, R.~Dermisek, J.~F.~Gunion and N.~Weiner,
  Ann.\ Rev.\ Nucl.\ Part.\ Sci.\  {\bf 58}, 75 (2008)
  [arXiv:0801.4554 [hep-ph]].


\bibitem{Schuster:2005py}
  P.~C.~Schuster and N.~Toro,
  arXiv:hep-ph/0512189.
  
  
\bibitem{Dermisek:2006wr}
  R.~Dermisek and J.~F.~Gunion,
  Phys.\ Rev.\  D {\bf 75}, 075019 (2007)
  [arXiv:hep-ph/0611142].
  
  
  
\bibitem{Dermisek:2005gg}
  R.~Dermisek and J.~F.~Gunion,
  Phys.\ Rev.\  D {\bf 73}, 111701 (2006)
  [arXiv:hep-ph/0510322].
  
  
\bibitem{Bellazzini:2009xt}
  B.~Bellazzini, C.~Csaki, A.~Falkowski and A.~Weiler,
  arXiv:0906.3026 [hep-ph].
  
  
  
\bibitem{Dermisek:2008id}
  R.~Dermisek,
  arXiv:0806.0847 [hep-ph].
  
\bibitem{:2004qh}
    [LEP Collaborations],
  arXiv:hep-ex/0412015.
  
  
\bibitem{Dermisek:2008dq}
  R.~Dermisek,
  arXiv:0807.2135 [hep-ph].
  
\bibitem{Ellis:1988er}
  J.~R.~Ellis, J.~F.~Gunion, H.~E.~Haber, L.~Roszkowski and F.~Zwirner,
  Phys.\ Rev.\  D {\bf 39}, 844 (1989).  
  
  
\bibitem{Dermisek:2008uu}
  R.~Dermisek and J.~F.~Gunion,
  Phys.\ Rev.\  D {\bf 79}, 055014 (2009)
  [arXiv:0811.3537 [hep-ph]].
  
  
  
\bibitem{Dermisek:2006py}
  R.~Dermisek, J.~F.~Gunion and B.~McElrath,
  Phys.\ Rev.\  D {\bf 76}, 051105 (2007)
  [arXiv:hep-ph/0612031].

    
  \bibitem{superB-talk}
  R.~Dermisek, talk given at Inaugural meeting for upgraded Belle proto-collaboration, KEK, Japan, 3-4 July, 2008.
  
  
  
  
\bibitem{Feng:1999mn}
  J.~L.~Feng, K.~T.~Matchev and T.~Moroi,
  Phys.\ Rev.\ Lett.\  {\bf 84}, 2322 (2000)
  [arXiv:hep-ph/9908309].

\bibitem{Bagger:1999sy}
  J.~A.~Bagger, J.~L.~Feng, N.~Polonsky and R.~J.~Zhang,
  Phys.\ Lett.\  B {\bf 473}, 264 (2000)
  [arXiv:hep-ph/9911255].
  
  
\bibitem{Blazek:2001sb}
  T.~Blazek, R.~Dermisek and S.~Raby,
  Phys.\ Rev.\ Lett.\  {\bf 88}, 111804 (2002)
  [arXiv:hep-ph/0107097].
  
\bibitem{Blazek:2002ta}
  T.~Blazek, R.~Dermisek and S.~Raby,
  Phys.\ Rev.\  D {\bf 65}, 115004 (2002)
  [arXiv:hep-ph/0201081].
  
  
\bibitem{Dermisek:2000hr}
  R.~Dermisek, A.~Mafi and S.~Raby,
  Phys.\ Rev.\  D {\bf 63}, 035001 (2001)
  [arXiv:hep-ph/0007213].
  
  
\bibitem{Choi:2005uz}
  K.~Choi, K.~S.~Jeong and K.~i.~Okumura,
  JHEP {\bf 0509}, 039 (2005)
  [arXiv:hep-ph/0504037].
  
\bibitem{Choi:2005hd}
  K.~Choi, K.~S.~Jeong, T.~Kobayashi and K.~i.~Okumura,
  Phys.\ Lett.\  B {\bf 633}, 355 (2006)
  [arXiv:hep-ph/0508029].

\bibitem{Kitano:2005wc}
  R.~Kitano and Y.~Nomura,
  Phys.\ Lett.\  B {\bf 631}, 58 (2005)
  [arXiv:hep-ph/0509039].
  
  
\bibitem{Lebedev:2005ge}
  O.~Lebedev, H.~P.~Nilles and M.~Ratz,
  arXiv:hep-ph/0511320.
  

  
\bibitem{Dermisek:2006qj}
  R.~Dermisek, H.~D.~Kim and I.~W.~Kim,
  JHEP {\bf 0610}, 001 (2006)
  [arXiv:hep-ph/0607169].
  
\bibitem{Bae:2007pa}
  K.~J.~Bae, R.~Dermisek, H.~D.~Kim and I.~W.~Kim,
  JCAP {\bf 0708}, 014 (2007)
  [arXiv:hep-ph/0702041].
  
  
\bibitem{Dermisek:2007qi}
  R.~Dermisek, H.~Verlinde and L.~T.~Wang,
  arXiv:0711.3211 [hep-ph].
  
\bibitem{Randall:1998uk}
  L.~Randall and R.~Sundrum,
  Nucl.\ Phys.\  B {\bf 557}, 79 (1999)
  [arXiv:hep-th/9810155].

\bibitem{Giudice:1998xp}
  G.~F.~Giudice, M.~A.~Luty, H.~Murayama and R.~Rattazzi,
  JHEP {\bf 9812}, 027 (1998)
  [arXiv:hep-ph/9810442].
  
\bibitem{Langacker:2007ac}
  P.~Langacker, G.~Paz, L.~T.~Wang and I.~Yavin,
  Phys.\ Rev.\ Lett.\  {\bf 100}, 041802 (2008)
  [arXiv:0710.1632 [hep-ph]].
  
\bibitem{Baer:2002by}
  H.~Baer, C.~Balazs, A.~Belyaev, R.~Dermisek, A.~Mafi and A.~Mustafayev,
  JHEP {\bf 0205}, 061 (2002)
  [arXiv:hep-ph/0204108].
  
  
\bibitem{Balazs:2003mm}
  C.~Balazs and R.~Dermisek,
  JHEP {\bf 0306}, 024 (2003)
  [arXiv:hep-ph/0303161].
  
 
   
  
    

\end{thebibliography}
\end{document}